\documentclass[aps,amssymb,showkeys]{revtex4}
\usepackage{graphicx}
\usepackage[dvips, bookmarks, colorlinks=true, plainpages = false, citecolor = red, urlcolor = blue, filecolor = blue]{hyperref}
\def\G{\Gamma}

\newcommand{\be}{\begin{equation}}
\newcommand{\ee}{\end{equation}}
\newcommand{\bea}{\begin{eqnarray}}
\newcommand{\eea}{\end{eqnarray}}
\newcommand{\ba}[1]{\begin{array}{#1}}
\newcommand{\ea}{\end{array}}

\newcommand{\z}{\zeta}

\begin{document}

\title{Exact factorization of correlation functions in 2-D critical percolation}

\author{Jacob J. H. Simmons$^+$}
\email{j.simmons1@physics.ox.ac.uk}
 \author{Peter Kleban}
\email{kleban@maine.edu}
 \affiliation{LASST and Department of Physics \& Astronomy,
University of Maine, Orono, ME 04469, USA 
\\$^+$current address: Rudolf Peierls Centre for Theoretical Physics, 1 Keble Road, Oxford OX1 3NP, UK}

\author{Robert M. Ziff}
\email{rziff@umich.edu}
\affiliation{MCTP and Department of Chemical Engineering, University of Michigan, 
Ann Arbor, MI 48109-2136 USA}

\date{\today}
\begin{abstract}
By use of conformal field theory, we discover several exact factorizations of higher-order density correlation functions in critical two-dimensional percolation.  Our formulas are valid  in the upper half-plane, or any conformally equivalent region.  We find excellent agreement of our results  with high-precision computer simulations.  There are indications that our formulas hold more generally.
  
\end{abstract}
\keywords{correlation functions, factorization, percolation}
\maketitle

\section{Introduction} \label{intro}
Correlation functions play an important role in the theory of fluids at thermal equilibrium, appearing in expressions for many experimental as well as theoretical quantities.  Higher-order correlations, i.e.\ correlations of quantities such as the density $\rho(x)$ at several points, e.g.\   $\langle \rho(x_1)\rho(x_2)\rho(x_3)...\rangle$, where the brackets denote a thermal average, occur in many contexts.   Calculating such quantities is therefore a central goal of the theory of fluids.  This is, however, an especially challenging problem, and many approaches have been proposed (for a review, see \cite{St}).  One idea is to factorize the higher-order correlations in terms of lower-order correlations (with fewer points).  In this article, we consider percolation in two dimensions at the percolation point in the upper half-plane (or any simply connected region).  In this case, by use of conformal field theory, we are able to exhibit several exact formulas in which three-point correlations factorize in terms of two-point correlations or correlations involving one point and an interval.  There are no similar exact results in the theory of fluids, to our knowledge.

Percolation in two-dimensional systems  has a long history, and has been examined by a very wide variety of methods, especially at the percolation point (see \cite{KSZ} for some representative references).  In this paper we report the results of calculations of correlation functions of the density in the upper half-plane.  Our formulas follow from conformal field theory \cite{BPZ,JCedge}, which is applicable to critical 2-D percolation in the continuum limit.

We focus on the density, defined as  the number of samples for which a site belongs to clusters satisfying some specified boundary condition (such as clusters touching certain parts of the boundary)
divided by the total number of samples $N$,
in the limit $N \to \infty$.   The density is an interesting and also practically important universal feature of percolation at the critical point.  Note that the density at a point $z$ of clusters which touch specified parts of the boundary is proportional to the probability of finding a cluster that connects those parts of the boundary with a small region around the point $z$.  Thus density correlation functions may also be viewed as probabilities of configurations with certain specified connections.

 In conformal field theory, operators in the bulk (except the unit operator) are defined so that their expectation valued vanishes, e.g.\ $\langle \psi(z) \rangle=0$.  Hence  the density at $z$ calculated below is  {\it subtracted}.  In particular, it will vanish when $z$ is sufficiently far from the other points or intervals with which it connects.

In a recent Letter \cite{KSZ}, we considered the problem of clusters simultaneously touching one or  two intervals on the boundary of a system, and also considered cases where those intervals shrink to points (anchor points).  In \cite{KSZ} we exhibited factorization in one particular case, demonstrated a relation to two-dimensional  electrostatics, and highlighted the universality of percolation densities. (In particular, we pointed out that by conformal covariance, the factorization is valid in any simply connected domain.)  In this paper, we find  more factorizations, and also give explicit expressions for the coefficients in the factorization formulas.

In addition, we confirm our theoretical results via numerical simulations to a high degree of accuracy. One case confirms numerical results in \cite{KSZ} using a different realization of percolation; simulations of our new predictions are also included.

In section \ref{der}, we first give the derivation of the factorization results, and then compare them with computer simulations.  Section \ref{dis} includes a few concluding remarks and discussion.

\section{predictions and simulations} \label{der}

This section begins by presenting the derivation of our formulas, then compares them with computer simulations. 

First we recapitulate the derivation of the factorization presented in  \cite{KSZ}.  Employing conformal field theory, applied to percolation, one may identify three operators of interest.   The boundary operator $\phi_{1,2}(x)$, with conformal dimension $h_{1,2}=0$, changes the boundary conditions from fixed to free at a point $x$ on the boundary (here, the real axis).  This operator appears in the field theory limit of a percolation system at a boundary point  between a segment where all the sites are occupied (or all empty) and one where the sites are unconstrained. Similarly, the boundary operator $\phi(x) := \phi_{1,3}(x)$, with conformal dimension $h_{\phi}=1/3$, anchors a cluster at  a point $x$ on a free boundary segment; this operator appears along a free boundary at a point where a cluster touches the boundary.  Finally, the ``magnetization" operator $\psi(z) := \phi_{3/2,3/2}(z)$, with conformal dimension $h_{\psi}=5/96$, measures the density of clusters at a point $z$ in the upper-half plane.  It is the field theory limit of the corresponding lattice quantity. Correlations involving these operators are proportional to the probability of finding a cluster connecting the various points, or intervals between boundary points, that they define.  

The notation in the following formulas omits, for brevity and clarity, various constants of proportionality.  Some of these are universal (independent of the particular realization of percolation), while others are not.  The latter type includes two kinds: constants multiplying the conformal operators, which are associated with the particular realization of the operator  for the system of interest, and constants specifying the dimension of the small regions with which the clusters are conditioned to connect. Our final results are homogeneous in operators and dimensional constants, so ignoring these constants makes no difference. The remaining (universal) constants are evaluated directly by taking an appropriate limit.

Consider the  probability ${\cal P}(x_1,x_2)$ of a cluster in the upper half-plane (as are all the clusters considered herein) that connects  the points $x_1$ and $x_2$ on the real axis. This (for $x_1 < x_2$) is 
\be \label{x1andx2}
{\cal P}(x_1,x_2) \propto \langle \phi(x_1) \phi(x_2) \rangle  \propto \Big( \frac{1}{x_2-x_1}\Big)^{2/3} \;.
\ee
This result follows from Cardy's formula for the crossing probability \cite{Cardy92} in the appropriate limit.

The probability ${\cal P}(z)$ of a cluster connecting  any point on the boundary with  a point $z=x+iy$ in the upper half-plane is simply 
\be \label{zandzb}
{\cal P}(z) \propto \langle \psi(z) \psi({\bar z}) \rangle \propto \Big(\frac{1}{y}\Big)^{5/48} \;,
\ee
consistent with the considerations in  \cite{FdG}.  Here,  the point ${\bar z}$ appears because because half-plane correlation functions  are given by full-plane correlators (which the expectation value denotes) with ``image" operators \cite{JCedge}.

The probability ${\cal P}(x_1,z)$ of a cluster constrained to touch  the (boundary) point $x_1$ and a point $z=x+iy$ in the upper-half plane is similarly given by a three-point correlation function,
\be \label{xandz}
{\cal P}(x_1,z) \propto \langle \phi(x_1) \psi(z) \psi({\bar z}) \rangle \propto  y^{11/48}\frac{1}{|z-x_1|^{2/3}} \;.
\ee

\begin{figure}
\begin{center}
\includegraphics[width=3in]{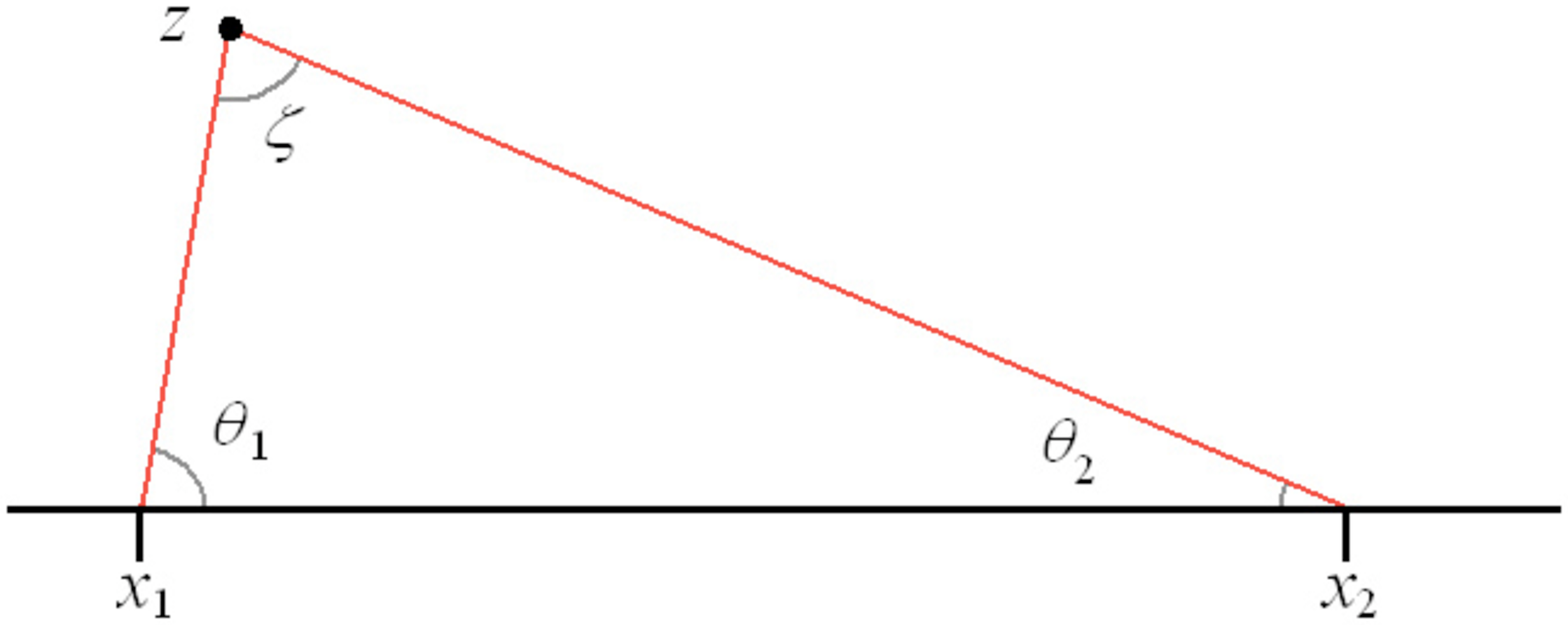}
\caption{(Color online).} The geometry used to define $\zeta$. \label{zfig}
\end{center}
\end{figure}

We now consider two more complicated objects. The first is  the probability ${\cal P}(x_1,x_2,z)$ of a cluster touching two boundary points $x_1$ and $x_2$ as well as a point $z$ in the upper-half plane. This is given by a four-point correlation function,
\be \label{x1x2z}
{\cal P}(x_1,x_2,z) \propto \langle\phi(x_1) \phi(x_2) \psi(z) \psi(\bar z)\rangle=(x_2-x_1)^{-2/3} y^{-5/48} F(\eta),
\ee
where the cross-ratio 
\be \label{edef}
\eta=\frac{(z-x_2)(\bar z-x_1)}{(\bar z-x_2)(z-x_1)} \;,
\ee
(this form is slightly different from, but equivalent to, the expression in \cite{KSZ}).  It is convenient, in what follows, to express the cross-ratio in terms of the angle $\z$ (see Fig. \ref{zfig})
\be \label{zdef}
\eta=e^{-2i(\theta_1+\theta_2)}=e^{2i\z}.
\ee

Since $\phi$ is a level-three operator, $F(\eta)$ satisfies a third-order differential equation.  The appropriate solution may be identified via physical arguments in the limit $x_1 \to x_2$ \cite{KSZ}.  In terms of $\z$, it is 
\be \label{2ptsol}
F(\z)={\rm sin}^{1/3}(\z).
\ee
Next, combining (\ref{x1andx2}) and (\ref{xandz}-\ref{2ptsol}) one finds by simple algebra that 
\be \label{fctr1}
{\cal P}(x_1,x_2,z) = C_1 \sqrt{{\cal P}(x_1,x_2)\;{\cal P}(x_1,z)\;{\cal P}(x_2,z)} \;.
\ee
Here, the constant $C_1$ may be evaluated by taking $x_1 \to x_2$, so that the lhs of (\ref{fctr1}) becomes a three-point function, with $C_1=C_{222}$  a (boundary) operator product expansion coefficient \cite{KSZ}.  This is evaluated in \cite{SK}, giving
\be \label{coeff1}
C_1 = \frac{2^{7/2}\;\pi^{5/2}}{3^{3/4}\;\G(1/3)^{9/2}} \;.
\ee
One finds $C_1=1.0299268 \ldots$.  In \cite{KSZ} we report simulation results verifying (\ref{fctr1}). For bond percolation on the square lattice we find $C_1=1.030 \pm 0.001$, in excellent agreement with (\ref{coeff1}).  Simulation also shows that (\ref{fctr1})  applies, with the same value of $C_1$, when one or both of the points $x_1,\;x_2$ is moved off the boundary  \cite{KSZ}.  However in this case the factorization only holds asymptotically, when the points are far apart compared to the distance to the edge. 

We emphasize that (\ref{fctr1}), along with (\ref{fctr2b}), (\ref{fctr3}) and (\ref{fctr4}) below, are both exact and universal.  Furthermore, they (by a conformal mapping) also hold in any simply-connected region, with the same proportionality constants.  In a different geometry, the functions in (\ref{fctr1}), (\ref{fctr2b}), (\ref{fctr3}) or (\ref{fctr4}) will change, but the relation remains, with the same proportionality constant.

Note that (\ref{fctr1}) resembles the Kirkwood superposition approximation \cite{K}, with the difference that here there is a square root and a coefficient on the rhs.  Further, the Kirkwood formula is  apparently only exact asymptotically, by contrast to (\ref{fctr1}), which is both exact and universal. 

Now we examine the probability ${\cal P}((x_1,x_2),z)$  \Big(${\cal P}\big(\overline{(x_1,x_2)},z\big)$\Big) of a cluster touching the boundary on (outside) the {\it interval} $(x_1,x_2)$ as well as a point $z$ in the upper-half plane. Both these quantities are given by the correlator
\be \label{12intz}
\langle\phi_{1,2}(x_1) \phi_{1,2}(x_2) \psi(z) \psi(\bar z)\rangle= y^{-5/48} G(\z) \;.
\ee
Since $\phi_{1,2}$ is a level-two operator, $G$ satisfies a second-order differential equation.  The solution corresponding to ${\cal P}((x_1,x_2),z)$ \cite{KSZ} may be written as
\be \label{intz}
{\cal P}((x_1,x_2),z) \propto y^{-5/48} {\rm sin}^{1/3}(\z/2) \;;
\ee
and it is straightforward to verify that
\be \label{bintz}
{\cal P}\Big(\overline{(x_1,x_2)},z\Big) \propto y^{-5/48} {\rm cos}^{1/3}(\z/2) \;.
\ee
The identity ${\rm sin}(\z)=2\;{\rm sin}(\z/2){\rm cos}(\z/2)$ then immediately implies
\be \label{fctr2a}
{\cal P}(x_1,x_2,z)\;{\cal P}(z) \propto {\cal P}(x_1,x_2)\;{\cal P}((x_1,x_2),z)\;{\cal P}\Big(\overline{(x_1,x_2)},z\Big) \;.
\ee

We can evaluate the constant in (\ref{fctr2a}) by taking the limit $x_1 \to x_2$ (the same procedure used to evaluate $C_1$ in (\ref{fctr1})).  The leading term gives
\be \label{fctr2b}
{\cal P}(x_1,x_2,z)\;{\cal P}(z) = C_2 \;{\cal P}(x_1,x_2)\;{\cal P}((x_1,x_2),z)\; {\cal P}\Big(\overline{(x_1,x_2)},z\Big) \;,
\ee
with the universal constant $C_2$ equal to the ratio of the boundary operator product expansion coefficients $C_{2 2 2}$, given above, to $C_{1 1 2}$, given below. Specifically, $C_2 = C_{222}/C_{112}$, so that
\be \label{coeff2}
C_2 = \frac{8 \pi^2}{3}\frac{1}{\G(1/3)^3} \;,
\ee
with $C_2 =1.36893 \ldots$.  Note that although (\ref{fctr2b}) includes  correlation functions involving specified intervals, which are perhaps more complicated than the correlation functions in (\ref{fctr1}), there is no square root.

Next, for completeness, we present two  factorized expressions that follow from the above, but have different forms and certain new features. First, one can eliminate ${\cal P}(x_1,x_2,z)$ between (\ref{fctr1}) and (\ref{fctr2b}).  This gives
\be \label{fctr3}
{\cal P}((x_1,x_2),z)\; {\cal P}\Big(\overline{(x_1,x_2)},z\Big) \;\sqrt{{\cal P}(x_1,x_2)} = C_3 \;{\cal P}(z)\; \sqrt{{\cal P}(x_1,z)\;{\cal P}(x_2,z)} \;,
\ee
with $C_3=C_{112}$. Thus (see \cite{SK})
\be \label{coeff3}
C_3=\frac{2^{1/2} \; 3^{1/4}\;\pi^{1/2}}{\G(1/3)^{3/2}},
\ee
and $C_3 =  0.752360738\ldots$.  Finally,  multiplying (\ref{fctr1}) by (\ref{fctr3}) (or dividing the square of (\ref{fctr1}) by (\ref{fctr2b})) gives
\be \label{fctr4}
{\cal P}((x_1,x_2),z)\; {\cal P}\Big(\overline{(x_1,x_2)},z\Big) \;{\cal P}(x_1,x_2,z) = C_4 \;{\cal P}(z)\; {\cal P}(x_1,z)\;{\cal P}(x_2,z) \;,
\ee
with  $C_4=C_1\;C_3$, so that
\be\label{coeff4}
C_4=\frac{2^4\;\pi^3}{3^{1/2}\;\G(1/3)^{6}} \;,
\ee
and thus $C_4 = 0.7748764775 \ldots$.  Equation (\ref{fctr4}) is ``homogeneous in averages", as discussed below.

\begin{figure}
\begin{center}
\includegraphics[width=5in,height=1.5in]{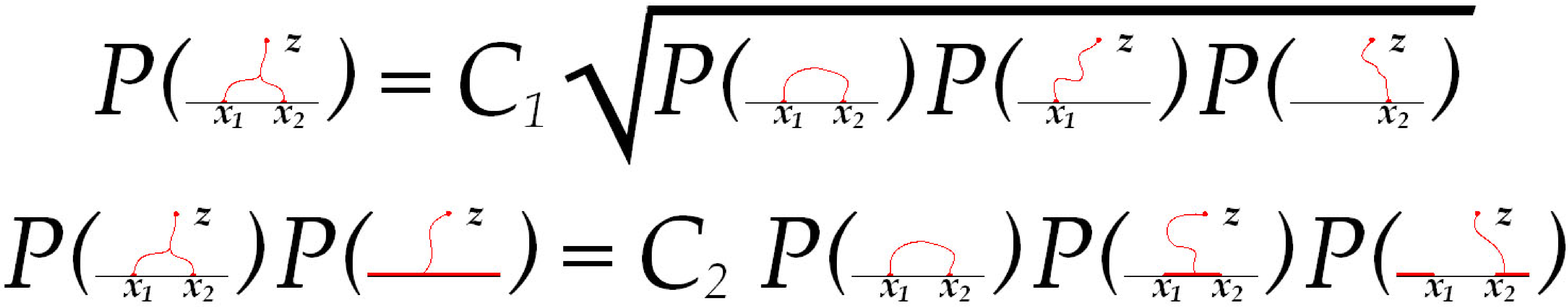}
\caption{(Color online).} Diagrammatic illustrations of (\ref{fctr1}) and (\ref{fctr2b}). \label{diags}
\end{center}
\end{figure}

Figure \ref{diags} illustrates (\ref{fctr1}) and (\ref{fctr2b}) diagrammatically.

\begin{figure}
\begin{center}
\includegraphics[width=5in]{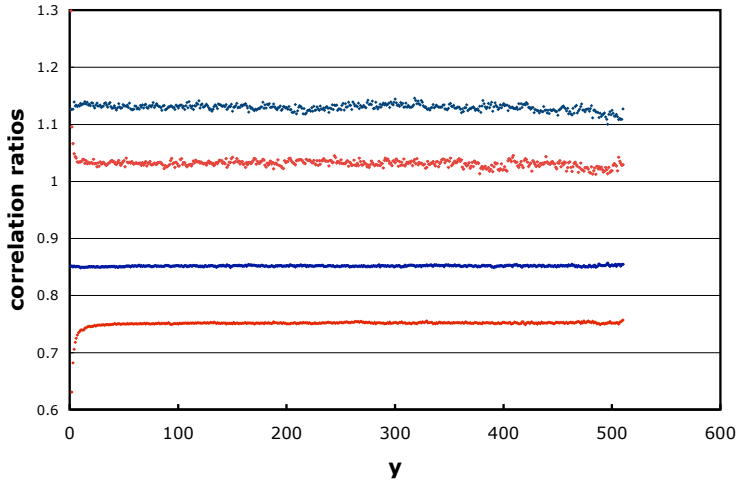}
\caption{(Color online).} Numerical results for correlation ratios predicted to be $C_1 = 1.0299...$
(upper two curves) and $C_3 = 0.75236...$ (lower two curves), for a
system of $510 \times 510$ sites,  for anchors or endpoints at 
$x_1 = 192$, $y = 1$, and $x_1 = 320$, $y = 1$.  In each pair of
curves, the upper (blue online) is along $x = 256$ (the centerline),
shifted up by $0.1$, while the lower (red online) represent the data
for $x = 192$, touching the point $x_1$.  \label{nums}
\end{center}
\end{figure}

We have simulated (\ref{fctr1}) and (\ref{fctr2b}) (note that \cite{KSZ} includes other numerical results for (\ref{fctr1})).  We used  site percolation on a square lattice of size $510 \times 510$, at $p_c = 0.5927463$, and $5 \times 10^6$ samples.
The boundary sites, chosen as $(x,y) = (x_1=192, 1)$ and $(x_2=320, 1)$  (i.e. $3/8$ and $5/8$ of  the way across bottom edge of the square), defined the interval $(x_1,x_2)$ (and its complement $\overline{(x_1,x_2)}$).  We considered each site $z$ in the entire lattice, and determined
which of the various boundary points and intervals that it connects with.  The fraction of samples in which the two interval boundary points were connected together was found to be
${\cal P}(x_1, x_2) = 0.0177522$. Figure \ref{nums} shows the ratio  ${\cal P}(x_1,x_2,z) / \sqrt{{\cal P}(x_1,x_2)\;{\cal P}(x_1,z)\;{\cal P}(x_2,z)}$, which is predicted by (\ref{fctr1}) to equal
$C_1 =C_{222}$, and also the ratio ${\cal P}((x_1,x_2),z)\; {\cal P}\Big(\overline{(x_1,x_2)},z\Big) \;\sqrt{{\cal P}(x_1,x_2)}/{\cal P}(z)\; \sqrt{{\cal P}(x_1,z)\;{\cal P}(x_2,z)}$  predicted to be $C_3= C_{112}$ by (\ref{fctr3}).  These quantities are shown along
the line $x = 192$, $0 < y < 512$, which includes the
 end-of-interval point $(x_1,1)$, and the vertical centerline  $x = 256$, $0 < y < 511$.
Clearly, the results are consistent with a constant for both quantities, except for deviations
a few lattice spacings away from the end point at $(x_1,1)$.  Similar finite-size effects
about the anchor points were
seen for the first ratio for the bond percolation case in \cite{KSZ}.   The
data for the first ratio  is less smooth than
that for the second, because the former depends upon ${\cal P}(x_1,x_2,z)$, which is a rarer event and more subject to fluctuations than the other quantities.

For the first ratio, we find $C_1=C_{222}=1.030 \pm 0.001$, in excellent agreement with the predicted value $C_1=1.0299\ldots$, and identical to the value found in  \cite{KSZ} using bond percolation on the square lattice.  For the second ratio, we find
$C_3=C_{112} =0.7529 \pm 0.001$, also in excellent agreement with the theoretical value $C_3 =  0.752360738\ldots$  These values were determined by averaging the point $z$ over all points on the lattice.

\section{discussion} \label{dis}
Any of our factorization results, namely (\ref{fctr1}),  (\ref{fctr2b}), (\ref{fctr3}) or (\ref{fctr4}), when written as expectation values of conformal operators, as mentioned, is necessarily ``homogeneous in operators", i.e.\ each distinct operator either appears to the same power on each side of the equation, or pairs of operators are  replaced by a single operator and the appropriate operator product expansion coefficient.  If this were not so, a universal expression would not be possible.  

Now (\ref{fctr1}) may be regarded as a generalization of the three-point function of conformal field theory \cite{P}, in the case when all three operators are the same.  (It reduces to this when the point $z$ is on the boundary.)  The three-point result only requires covariance under the special conformal group, not the full machinery of conformal field theory. This may indicate that (\ref{fctr1}) is more generally valid.

Equation (\ref{fctr4}) is ``homogeneous in averages", i.e.\ the same number of brackets appears on each side.  This means that it may be verified numerically without overall normalization--one can use the raw data for the number of samples for each specified probability $\cal{P}$ without dividing by the total number of samples.

As remarked in \cite{KSZ}, preliminary calculations  and numerical data show that  factorization generalizing (\ref{fctr1}) holds for  Fortuin-Kastleyn clusters in the critical Potts models as well \cite{KSSZ}.   

It is interesting to understand the factorization in terms of two-dimensional electrostatics.  There are several ways to do this.  Defining $\varphi(x;z)=1/(z-x)$ as the generalized (complex) potential at $z$ of a unit dipole at $x$ expresses the potential of a dipole of strength $|p|$ in the direction ${\rm Arg}(p)$ as $p \; \varphi(x;z)$.  Now, establishing our factorization results involves writing ${\cal P}(x_1,x_2,z)$, ${\cal P}((x_1,x_2),z)$, or  ${\cal P}\big(\overline{(x_1,x_2)},z\big)$ in terms of simpler correlation functions.  The key algebraic step needed is expressible as
\be
{\rm sin}^2(\zeta) \propto |\varphi(x_1;z) \varphi(x_2;z)|^2  y^2 (x_1-x_2)^2 \propto \frac{ |\varphi(x_1;z) \; \varphi(x_2;z)|^2}{|\varphi(z_{\;};{\bar z_{\;}})\; \varphi(x_1;x_2)|^2}\;.
\ee
This may be used, for example, in conjunction with (\ref{zandzb}-\ref{2ptsol}), (\ref{intz}), and (\ref{bintz})  to establish  (\ref{fctr4}). (These manipulations do not give an expression for $C_4$, of course.)

Finally, one might wonder why the numerical accuracy of the conformal prediction is so high, especially at short distances.  In general, the field theory limit of an appropriate lattice quantity, for instance the order parameter, is given by a conformal primary field plus correction terms proportional to its descendant fields.  These descendants necessarily have dimensions larger than that of the primary, and hence give rise to terms in a given correlation function that die away more rapidly with distance than those due to the primary, but which may be substantial at short distances, even for very large lattices.  However, in critical percolation in two dimensions, previous numerical work on closely related quantities (see \cite{KSZ,Z}) has shown that such effects are very small.  The reason for this is, to our knowledge, not known, but it follows that the accuracy which we observe herein is not surprising.

 In conclusion, we have presented several new formulas for higher-order correlation functions applicable to critical percolation in two dimensions.  These have the property of exact factorization in terms of lower-order correlations or correlations involving intervals. Our predictions agree with the results of high-precision simulations.
 
 For the future, it might be possible, using perturbation theory, to find the corrections to our factorization results for $p \ne p_c$.  However the calculations required do not appear to be simple.
   
\section{Acknowledgments}
We thank J. Cardy and J. K. Percus for their comments.
This work was supported in part by the National Science Foundation under grants numbers  DMS-0553487 (RMZ); and DMR-0203589 and continuation grant DMR-0536927 (PK).



\begin{thebibliography}{99}

 \bibitem{St}
 G. Stell, in {\it The equilibrium theory of classical fluids},  H. L. Frisch and J. L. Lebowitz, eds. (New York: W. A. Benjamin) 1964.

 \bibitem{KSZ}
  Peter Kleban,  Jacob J. H.  Simmons, and Robert M. Ziff,  {\it Anchored critical percolation clusters and 2D electrostatics},
 Phys. Rev. Lett.
  {\bf 97} 115702 (2006) \href{http://arxiv.org/abs/cond-mat/0605120}{[arXiv: cond-mat/0605120]}.
  
\bibitem{BPZ}
A. A. Belavin, A. M. Polyakov, and A. B. Zamolodchikov, {\it Infinite conformal symmetry in two-dimensional quantum field theory}, Nucl. Phys. {\bf B241}, 333-380 (1984).
  
\bibitem{JCedge}
J. L. Cardy, {\it Conformal invariance and surface critical behavior}, Nucl. Phys. {\bf B240 [FS 12]}, 514-522 (1984). 
  
\bibitem{Cardy92}
J. L. Cardy, {\it Critical percolation in finite geometries}, J. Phys. A
{\bf 25}  L201-206 (1992) \href{http://arxiv.org/abs/hep-th/9111026}{[arXiv: hep-th/9111026]}.

\bibitem{FdG}
M. E. Fisher  and P. G. de Gennes, {\it Ph{\' e}nom{\` e}nes aux parois dans un m{\' e}lange binaire critique:
  physique des collo{\" i}des },  C. R. Acad. Sci., Paris B {\bf 287} 207 (1978).
 \bibitem{SK}
Jacob J. H.  Simmons and  Peter Kleban,  {\it in preparation}.

\bibitem{K}
J. G. Kirkwood, {\it Statistical mechanics of fluid mixtures}, J. Chem. Phys. {\bf3} 300-313 (1935).

\bibitem{P}
 A. M. Polyakov, Plsma ZhETP {\bf 12} 538 (1970) [JETP Lett {\bf 12} 381 (1970)].

 \bibitem{KSSZ}
  Peter Kleban,  Jacob J. H.  Simmons, Thomas Stone and Robert M. Ziff,  {\it in preparation}.
  
  \bibitem{Z}
R. M. Ziff, {\it Effective boundary extrapolation length to account for finite-size effects in the percolation crossing function}, Phys. Rev. E {\bf 54}, 2547-2554 (1996).


\end{thebibliography}
\end{document}